\documentclass[twocolumn,trackchanges]{aastex61}
\usepackage{ulem}

\submitjournal{ApJ}

\shorttitle{HD111520: Multi-band}
\shortauthors{Crotts et al.}

\begin{document}

\title{A Multi-Wavelength Study of the Highly Asymmetrical Debris Disk Around HD 111520}
	
\correspondingauthor{Katie Crotts}
\email{ktcrotts@uvic.ca}

\author[0000-0003-4909-256X]{Katie A. Crotts}
\affil{Physics \& Astronomy Department, University of Victoria, 3800 Finnerty Rd. Victoria, BC, V8P 5C2}

\author[0000-0012-3245-1234]{Zachary H. Draper}
\affil{Physics \& Astronomy Department, University of Victoria, 3800 Finnerty Rd. Victoria, BC, V8P 5C2}
\affil{Herzberg Astronomy and Astrophysics, National Research Council of Canada, 5071 West Saanich Rd., Victoria, BC V9E 2E7, Canada}

\author[0000-0003-3017-9577]{Brenda C. Matthews}
\affil{Herzberg Astronomy and Astrophysics, National Research Council of Canada, 5071 West Saanich Rd., Victoria, BC V9E 2E7, Canada}

\author[0000-0002-5092-6464]{Gaspard Duch\^{e}ne}
\affiliation{Astronomy Department, University of California, Berkeley, CA 94720, USA}
\affiliation{Universit\'{e} Grenoble Alpes/CNRS, Institut de Plan\'{e}tologie et d'Astrophysique de Grenoble, 38000 Grenoble, France}

\author[0000-0002-0792-3719]{Thomas M. Esposito}
\affiliation{Astronomy Department, University of California, Berkeley, CA 94720, USA}
\affiliation{SETI Institute, Carl Sagan Center, 189 Bernardo Ave., Mountain View CA 94043, USA}

\author[0000-0003-1526-7587]{David Wilner}
\affiliation{Center for Astrophysics | Harvard \& Smithsonian, 60 Garden Street, Cambridge, MA 02138, USA}

\author{Johan Mazoyer}
\affiliation{LESIA, Observatoire de Paris, Universit\'{e} PSL, CNRS, Sorbonne Universit\'{e}, Universit\'{e} de Paris, 5 place Jules Janssen, F-92195 Meudon, France}

\author[0000-0001-5334-5107]{Deborah Padgett}
\affiliation{Jet Propulsion Laboratory, California Institute of Technology, 4800 Oak Grove Drive Pasadena, CA 91109 USA}

\author[0000-0002-6221-5360]{Paul Kalas}
\affiliation{Astronomy Department, University of California, Berkeley, CA 94720, USA}
\affiliation{SETI Institute, Carl Sagan Center, 189 Bernardo Ave., Mountain View CA 94043, USA}
\affiliation{Institute of Astrophysics, FORTH, GR-71110 Heraklion, Greece}

\author[0000-0002-2805-7338]{Karl Stapelfeldt}
\affiliation{Jet Propulsion Laboratory, California Institute of Technology, 4800 Oak Grove Drive Pasadena, CA 91109 USA}

\begin{abstract}

We observed the nearly edge-on debris disk system HD 111520 at $J$, $H$, \& $K1$ near infrared (NIR) bands using both the spectral and polarization modes of the Gemini Planet Imager (GPI). With these new observations, we have performed an empirical analysis in order to better understand the disk morphology and its highly asymmetrical nature. We find that the disk features a large brightness and radial asymmetry, most prominent at shorter wavelengths. We also find that the radial location of the peak polarized intensity differs on either side of the star by 11 AU, suggesting that the disk may be eccentric, although, such an eccentricity does not fully explain the large brightness and radial asymmetry observed. Observations of the disk halo with HST also show the disk to be warped at larger separations, with a bifurcation feature in the northwest, further suggesting that there may be a planet in this system creating an asymmetrical disk structure. Measuring the disk color shows that the brighter extension is bluer compared to the dimmer extension, suggesting that the two sides have different dust grain properties. This finding, along with the large brightness asymmetry, are consistent with the hypothesis that a giant impact occurred between two large bodies in the northern extension of the disk, although confirming this based on NIR observations alone is not feasible. Follow-up imaging with ALMA to resolve the asymmetry in the dust mass distribution is essential in order to confirm this scenario. 
\end{abstract}


\section{Introduction} \label{intro}
Debris disks are optically thin, dust disks around stars, which are generated through the perturbation of planetesimals in the system. This causes what's known as a ``collisional cascade" of larger bodies that produces micron to millimeter sized dust grains \citep{MW08,BM14,MH18}. As instrumentation has improved over the last decade, allowing for higher resolution imaging, debris disks have been found to harbor many types of asymmetries and structures such as gaps, rings, eccentricities, brightness asymmetries, and warps (\citealt{MH18} and references therein). While such features are often thought to be caused by planets, other mechanisms such as an interaction with the interstellar medium (ISM, \citealt{DJ09}) or giant impacts can shape debris disks as well. For example, an ISM interaction was originally thought to have formed the moth- and needle-like halos of the debris disks around HD 61005 and HD 11515 \citep{Schneider:2014aa,Rodigas:2012aa}, while the brightness asymmetry and CO clump in the SW extension of $\beta$ Pic can be explained by a recent massive impact \citep{WD14,DA15}. Additionally, the majority of resolved debris disk systems have no known planet companions to directly connect with the disk morphology, demonstrating the need for a more comprehensive understanding of how debris disks are perturbed.

Due to the response of the dust component of debris disks to dynamical perturbations, they are good indicators of a planet's stability, where dust belts can serve as indicators for planetary upheaval by large planetary perturbations or whether minor sculpting could be occurring \citep{TE16,SM17}. Through dynamical modelling, it has been shown that even a single 10 M$_{\oplus}$ planet on an eccentric orbit can produce many different debris disk morphologies and asymmetries \citep{Lee:2016aa}. In some cases disk structure can be directly traced to a known planet sculpting the disk such as in the case of $\beta$ Pic \citep{mouillet97,Dawson2011,DA15}. Another example of this is the asymmetrical debris disk HD 106906, which exhibits both a brightness asymmetry and eccentricity \citep{PK15,AL15,KC21}. Through empirical analysis, these asymmetries are shown to be likely due to the 11 M$_{Jup}$ planet in the system that orbits outside of the disk, and unlikely to be due to an ISM interaction \citep{KC21}. While both $\beta$ Pic and HD 106906 have known planetary companions, there are other highly asymmetrical debris disks in which no known planets exist. One example is the debris disk HD 160305, recently discovered in 2019, which harbors a large azimuthal brightness asymmetry accredited to either a hidden planet companion or a recent large impact \citep{perrot2019}. Studying highly perturbed debris disks can therefore offer insight into the stirring mechanisms of dust belts, and whether their morphologies were produced by planets in the system or through other phenomena. 

HD 111520 (HIP 62657) is located 108.1$\pm$0.2 pc from the Sun \citep{gaia21} and is a member of the Lower Centaurus Crux (LCC) group within the Scorpius-Centaurus Association \citep{dezeeuw99}. As part of the GPIES campaign \citep{Mac18,BM14,BM08}, which resolved 26 debris disks in polarized and total intensity \citep{TE20}, one of the systems observed was HD 111520. The debris disk was detected in the $H$ band and revealed a strongly asymmetric disk morphology from 0.3-1$''$, with a 2:1 brightness asymmetry and radial asymmetry measured between the two sides of the disk \citep{zd16b}. From Hubble Space Telescope (HST) observations, the system had previously been shown to have an asymmetric, ``needle"-like disk structure out to 6$''$ ($\sim$650 AU) relative to the central star, along with an even larger 5:1 brightness asymmetry and a bifurcation feature in the Northern extension \citep{DP15}. Such a strong asymmetry could be the result of two main scenarios, either the dust grain properties are significantly different between the two extensions, which changes the dust scattering efficiency, or the planetesimal belt itself is being perturbed by dynamical activity, such as that which could result from an unseen planetary companion.

Since HD 111520 is such an unusual system, we've obtained multi-wavelength GPI data to better characterize the disk morphology and further investigate its observed asymmetry. We have conducted an empirical analysis of our GPI data through measuring the disk structure in Section \ref{structure}, as well as the surface brightness profiles of each band and disk color in Section \ref{sbsection}. We then discuss possible explanations for the disk's asymmetries based on the results of our empirical analysis in Section \ref{discussion}.

\begin{table*}
	\centering
	\caption{\label{data_sum}Summary of the data used in this paper. The H Spec data is the same data from \cite{zd16b}, while the H Pol data (previously published in \citealt{TE20}) was a subsequent longer observation (582 s vs. 2840 s integration time). Here, N = the number of frames, t$_{int}$ = the total integration time in seconds, and $\Delta$PA = the total parallactic angle rotation in degrees.}
	\begin{tabular*}{\textwidth}{c @{\extracolsep{\fill}} cccccc}
	    \hline
	    \hline
		Band & Mode & Date & N & t$_{int}$ (s) & $\Delta$PA (\degr) & MASS Seeing (\arcsec) \\
		\hline
		J & Pol & 2016 Mar 26 & 58 & 3480 & 39.1 & 0.82\\
		J & Spec & 2016 Mar 27 & 51 & 3060 & 29.5 & 0.45\\
		H & Pol & 2016 Mar 18 & 26 & 2840 & 28.3 & 0.48\\
		H & Spec & 2015 Jul 02 & 41 & 2446 & 34.8 & 0.28\\
		K1 & Pol & 2016 Mar 28 & 36 & 2160 & 35.8 & 1.33\\
		K1 & Spec & 2016 Mar 24 & 30 & 1800 & 19.2 & 0.59\\
		\hline
		\hline
	\end{tabular*}
\end{table*}

\section{Observations and Data Reduction} \label{reduction}

The imaging of HD 111520 from GPI was collected over a range of nights, combining data sets from the GPIES survey in $H$ band ($\sim$1.65 $\mu$m; PI: Bruce Macintosh) and the Debris Disk Large and Long Program in $J$ and $K1$ bands ($\sim$1.24 $\mu$m and $\sim$2.05 $\mu$m; PI: Christine Chen). Observations were taken in both polarimetric and spectroscopic modes, with a field of view (FOV) of $2\farcs8$ by $2\farcs8$ and a pixel scale of 14.166$\pm$0.007 mas lenslet$^{-1}$ \citep{derosa15}. A summary of the observations can be seen in Table \ref{data_sum}. In general, the observations were scheduled to maximize field rotation as the source transited the meridian, with a total observing sequence of around an hour (including overheads) with a series of 60 second exposures to optimize PSF subtraction. The $H$-band polarization mode data set, previously published in \cite{TE20}, is a longer integration time than the data presented in \cite{zd16b}, therefore achieving a higher signal-to-noise ratio (S/N), comparable to the other bands.

The data reduction in general followed standard practices for GPI data employed by the GPI pipeline (\citealt{per14}, and references therein). The polarization mode (pol-mode) is observed as two orthogonal polarization states on the detector and is modulated by an achromatic half-wave plate (HWP) to four different orientations \citep{MP15}. The relative measurement of flux between them can later be used to derive the Stokes parameters in the field of view. The detector images are dark subtracted and `destriped' with a Fourier-filter to remove a standing wave pattern from microphonic noise \citep{PI14}. The flexure in the instrument is compensated using a cross-correlation algorithm to match the detector with the expected positions \citep{ZD14}. The extracted spot fluxes create the pol-mode data cube and can then be cleaned for bad pixels and systematic offsets using a modified double difference algorithm \citep{per14}. A geometric distortion correction in the FOV from astrometric standards are then used \citep{QK14}. The instrumental polarization is also removed based on measurements of the flux under the coronagraph \citep{MB15}. Additionally, the data are smoothed by a Gaussian with a FWHM of 1 pixel. The data were then flux calibrated by measuring the satellite spots produced by GPI's optics with elongated apertures (since the spots are smeared out by diffraction with wavelength in the broad band pol-mode) and compared to 2MASS JHK magnitudes of the star to calibrate the flux within the data cube \citep{LH15}. The spots are also used to determine the star center by using a Radon transform algorithm \citep{JW14,MP15}. The data set at multiple position angles and HWP orientations is then combined with a singular value decomposition on Mueller matrices to create a Stokes data cube. The values are converted to a radial Stokes convention to put the stellocentric polarized emission into a single Stokes mode ($Q_{\phi}$) \citep{HS06}.

In the case of the spectroscopic mode, again standard GPI pipeline data reduction practices are used \citep{per14}. Each of the individual dispersed light frames were dark subtracted, had bad pixels masked, and were `destriped' from microphonics \citep{PI14}. Wavelength calibrations are done using Ar lamp exposures in each respective band \citep{SW14}. Calibration sources are measured prior to the observing sequence, minimizing flexure offsets. For the $J$ and $H$ bands a standard box aperture method is used to extract the flux into a wavelength calibrated data cube \citep{JM14s}. In the $K1$ band, the satellite spots had weaker flux compared to the background, which interferes with the calibrations necessary after cube extraction. The microlens PSF extraction method was then used to extract the data from the detector \citep{ZD14,PI14}. The S/N improved enough for the pipeline to find the satellite spots in subsequent steps in at least the central wavelength slice of each $K1$-band IFS cube. Sky-subtraction at the 2D detector was attempted but this step appeared to introduce more noise. The sky subtraction was therefore left to PSF subtraction at later steps using pyKLIP \citep{JW15}. For bad spaxel mitigation, a bad pixel identifier and smoothing was applied similar to \cite{zd16b}. The satellite spots were then identified by centering on a high pass filtered image of the data. A best guess position for a slice with high S/N was identified by eye once for the whole sequence to aid the spot location algorithm. The unfiltered images then had their satellite spots extracted to measure the stellar spectrum convolved with the instruments response function in wavelength and time. The spectrum of each cube was calibrated by comparing the measured flux with the 2MASS magnitude for each respective band. A color correction was applied by comparing the measured spectrum to a real, atlas template spectrum \citep{AP98} for a star of HD 111520's spectral type (F5V). These two factors essentially measure the flux conversion factors for each cube so that the satellite spots have the same absolute flux and spectra as an F5V star. This calibrates the data in the cube to physically relevant units of Janskys. This algorithm was tested on a known white dwarf spectrum in \cite{JM14sb} and found to be within a 5\% flux error with on sky observations. The reduced data cubes are then run through the program, pyKLIP, which utilizes angular-differential-imaging and subtracts the stellar PSF to remove additional flux from the star. For a more in-depth description of this process, we refer to \cite{zd16b}. 

\begin{figure*}[ht!]
\centering
	\caption{\label{poliandtoti} \textbf{Top:} The total intensity (TI) detections of HD 111520, produced by pyKLIP. \textbf{Middle:} Polarized intensity (PI) of HD 111520. The polarized intensity is taken from the rotated stokes frame ($Q_{\phi}$) to isolate the astrophysical emission. \textbf{Bottom:} S/N of the polarized intensity detections measured by dividing noise maps from the $Q_{\phi}$ data (see Section \ref{reduction}). The disk is consistently dimmer on the SE (left) extension of the disk compared to the NW (right) extension of the disk. The circles represent the size of the focal plane mask ($0\farcs09$, $0\farcs12$, $0\farcs15$ for $J$, $H$, and $K1$ respectively), and the crosses represent the location of the star. For all data, East is left and North is up.}
	\includegraphics[width=\textwidth]{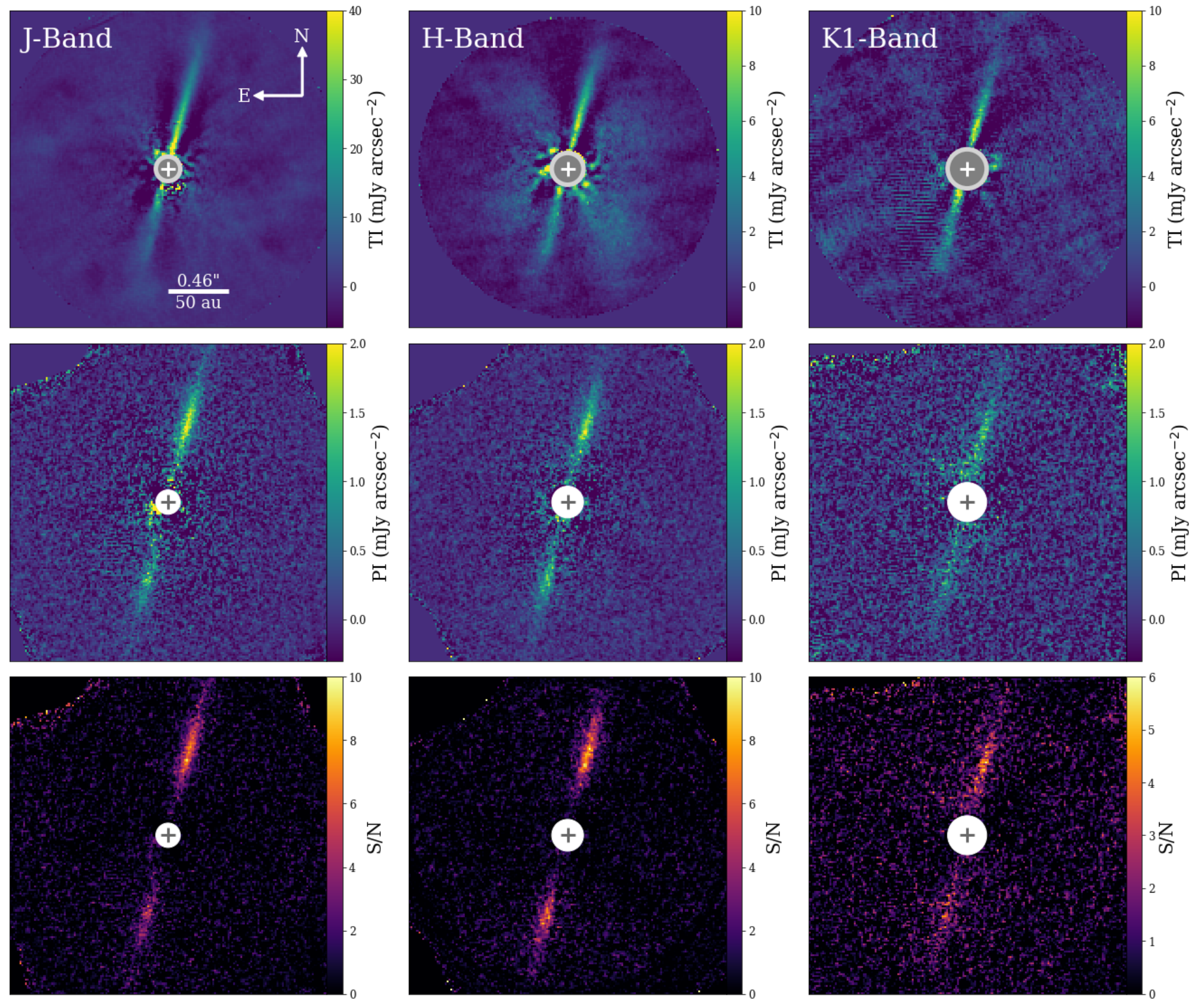}
\end{figure*}

The final images resolving the disk on both polarized and total intensity can be seen in Figure \ref{poliandtoti}. Additionally, we also include the S/N of our polarized intensity data, which are generated by dividing the $Q_{\phi}$ images by noise maps derived from the $U_{\phi}$ images. To create these noise maps, we use $U_{\phi}$ to calculate the standard deviation at each radius in 1-pixel wide stellocentric annuli. Here, we are assuming that $U_{\phi}$ contains no disk signal, which would be expected for an optically thin debris disk causing single scattering.

To better understand the prior HST results, we reprocessed the archival STIS data from GO-12998 (PI Padgett). HD 111520 was acquired behind the WEDGEA1.0 mask position and imaged in two consecutive orbits with two different roll angles separated by 32$\degr$. Each orbit accumulated five 419s integrations. PSF subtraction was accomplished by differencing the two calibrated and registered $\_$sx2.fits images from each other. The two PSF-subtracted images were rotated to north and the data were averaged.

\section{Observational Analysis}
To better understand the disk morphology and asymmetry, we perform an empirical analysis on our multi-wavelength GPI data. This includes measuring the vertical structure through the disk FWHM and the vertical offset from the star (i.e., the location of the disk spine), as well as measuring the surface brightness in each band. Characterizing the disk morphology allows us to probe the cause of the observed asymmetries, as different disk perturbers affect the disk differently. For example, interaction with a planet companion may cause structures in the disk geometry such as warps and eccentricities, whereas an interaction with the ISM or a giant impact may cause an asymmetry in the dust grain properties creating an asymmetrical disk color. These results will be further discussed in the context of possible disk perturbers in Section \ref{discussion}.

\subsection{Vertical Structure} \label{structure}
To measure the vertical structure, we fit a 1-D Gaussian profile to the vertical surface brightness profile at a series of radii from the star. As the disk structure should not vary significantly within our small wavelength range, we choose to focus this analysis mainly on the $H$ band, as it has the best S/N, but also perform the analysis on the $J$ and $K1$ band in order to compare. Additionally, we utilize the polarized intensity data, as we believe it most accurately represents the true disk structure, without the biases introduced by ADI-based PSF subtraction employed for the total intensity data. 

\begin{figure*}
	\caption{\label{vert} \textbf{Left:} The vertical width (FWHM) profile of HD 111520 as a function of separation from the star in each band. The grey horizontal dashed line represents the measured weighted FWHM, while the red line represents the intrinsic FWHM as measured in the $H$ band. The dark blue line represents the FWHM of the GPI $H$-band PSF. \textbf{Right:} The vertical offset profile as a function of the separation from the star in each band. The dashed grey line represents the best fitting narrow ring model for the $H$ band. For both profiles, we exclude measurements within $0\farcs3$ and greater than 1$''$ (with exception of the $J$ band) due to low signal-to-noise.}
	\includegraphics[width=1\textwidth]{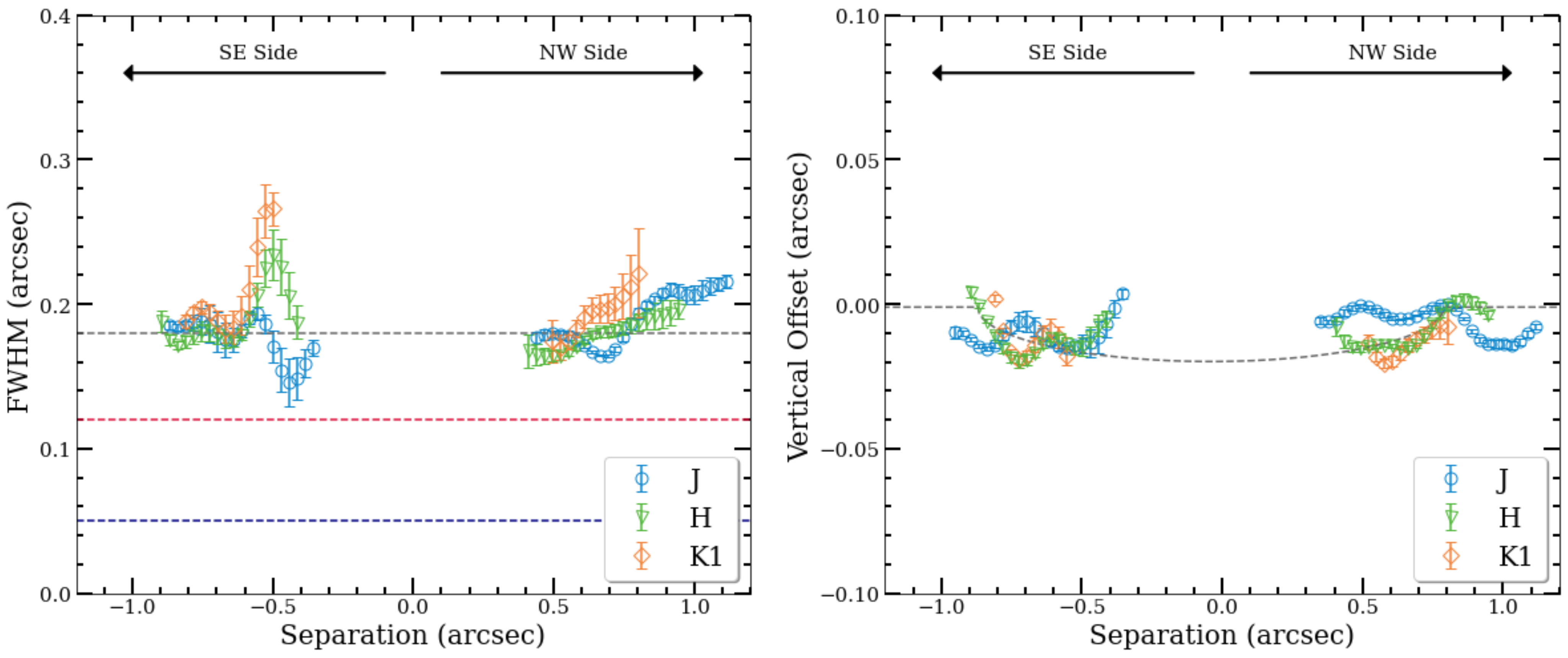}
\end{figure*}

We first prepared the data by rotating our data by 75$^{\circ}$ clockwise, similarly done in \cite{zd16b}. The data are then binned into $2 \times 2$ pixel bins, and additional smoothing using a Gaussian with $\sigma$=2 pixels ($0\farcs028$) is applied. This is done to increase the S/N and ensure a good fit to the data. Once the Gaussian profile is fit to vertical surface brightness slices along the entire disk using a non-linear least squares fit, the mean value and full-width-half-max (FWHM), along with 1$\sigma$ uncertainties of each Gaussian fit are extracted. The mean value (i.e., the center of the Gaussian) represents the disk spine, or more specifically, the vertical offset from a horizontal line passing through the star, whereas the FWHM represents the approximate vertical width of the disk. The resulting FWHM and vertical offset profiles for all three bands can be seen in Figure \ref{vert}. We are able to detect the emission in the $H$ band out to $\sim$1$''$ on either side of the disk. While the $K1$ band is also detected symmetrically (although only out to $0\farcs8$ given the lower S/N), the $J$ band does show a radial asymmetry where the SE extension is detected out to $\sim$1$''$, and the NW extension is detected to $\sim1\farcs1$, which corresponds to a difference of 11 AU. 

Taking the weighted mean of the $H$-band FWHM profile, we find the disk to have an approximate FWHM of $0\farcs18$. This is much greater than the GPI's instrumental $H$-band PSF FWHM of $0\farcs05$, showing that the disk is well resolved. However, this instrumental PSF and any smoothing/binning of the data must be taken into account in order to obtain an intrinsic measurement for the FWHM. This is done by subtracting in quadrature the FWHM of the instrumental PSF, as well as the FWHM of any smoothing applied, from the measured FWHM. Doing so, we obtain an intrinsic FWHM of 0.12$\pm$0.1$''$ for the $J$ and $H$ band, and 0.14$\pm$0.1$''$ for the $K1$ band (13-15 AU), which leads to a aspect ratio of $\sim$0.28 at 50 AU. This aspect ratio is intermediate between similar measurements at 50 AU for other near edge-on debris disks such as HD 32297 (0.17, \citealt{duchene20}) and HD 106906 (0.31, \citealt{KC21}), although these comparisons are purely from empirical estimates and not from a proper disk model. A positive trend can be seen in the NW extension in all three bands, showing that the vertical width increases with radial distance on this side, while the SE extension is flat past $0\farcs6$. This differs slightly from \cite{zd16b}, which only shows a positive trend in the NW extension past $0\farcs7$. Additionally, an enhancement in the vertical width can be seen in the SE at a separation of $\sim0\farcs5$, strongest in the $H$ and $K1$ bands, but also visible in the $J$ band. A similar feature was observed in \cite{zd16b}, although we find this enhancement to be $0\farcs1$ farther from the star. Given that the previous measurements were done with the total intensity $H$-band data, which are likely affected by PSF subtraction, our results using the polarized intensity are likely better to represent the true disk vertical width.

The vertical offset along the disk is small, with the largest offset being less than $0\farcs03$, due to the disk being highly inclined. However, a clear offset can be seen in each band in Figure \ref{vert}, showing that the disk inclination is not exactly 90$^{\circ}$. In all three bands, the majority of data points are negative, indicating that the front side of the disk lies to the west, contrary to what is observed in \cite{zd16b}. We are also able to clearly detect a vertical offset in the SE extension, whereas the previous measurements in \cite{zd16b} were unable to do so, showing the improvement that our polarized data have in measuring the vertical profile of the disk. 

To further constrain the disk geometry, we fit the vertical offset or disk spine with a narrow, inclined ring model, similarly used in \cite{duchene20}. We perform the fitting for all three bands, which will allow us to determine which parameters can be well constrained and which ones cannot. For the fitting procedure, MCMC via the python package \textit{emcee} \citep{df13} is utilized. Ring models are first generated from the equation of an ellipse with a given disk radii ($R_{d}$), x and y disk offset ($\delta_{x}$, offset along the major axis, and $\delta_{y}$, offset along the minor axis.), disk inclination ($i$), and position angle ($PA$) measured from East of North through rotating the data. Note that for the disk offsets, a negative value means a disk offset towards the left or down, while a positive value means a disk offset towards the right or up. These models are then compared to the data points using a $\chi^{2}$ function. The results from this fitting procedure for all three bands can be found in Table \ref{vert_sum}. 

Through this modelling, there are three parameters that we can better constrain: the inclination, $PA$ and $\delta_{y}$. As expected, we find HD 111520's debris disk to  have a very high inclination, no more than $2\fdg7$ away from completely edge-on given 3$\sigma$ uncertainties of the $PA$ in the $K1$ band. Our measurement of the $PA$ is also consistent with that measured in \cite{zd16b}, where we find it to lie between 165$^{\circ}$ and 166$^{\circ}$. We find no significant offset along the minor-axis ($\delta_{y}$), with an offset of $\lesssim0\farcs01$. Taking into consideration the uncertainty in the position of star, which is 0.05 pixels (or $\sim$0.7 mas) for GPI \citep{JW14}, makes this small offset negligible. Unfortunately, due to the high inclination of the disk, the disk radii, $R_{d}$, and offset along the major-axis, $\delta_{x}$, are too difficult to constrain from this type of modelling, and varies significantly between bands.

A possible warp is identified in \cite{zd16b}, where the SE extension was found to not align perfectly with the NW extension. However, such a warp is not seen in our vertical offset profiles. As a confirmation, we also check the vertical offset profiles for all three bands of the Spec data. Doing so, we find that this potential warp is also not present, showing that the warp feature is likely an artifact and a result of PSF subtraction in the H-band Spec data. This again shows that our higher S/N data greatly improves our measurements of the vertical profile and helps better constrain the disk's morphology and orientation.

\begin{table*}
	\centering
	\caption{\label{vert_sum}Measured properties of the vertical profile for each band. This includes the weighted intrinsic FWHM, along with the best fit inclined ring model parameters.}
	\begin{tabular*}{\textwidth}{c @{\extracolsep{\fill}} cccccc}
	    \hline
	    \hline
		Band & FWHM ($''$) & R$_{d}$ ($''$) & $\delta_{x}$ ($''$) & $\delta_{y}$ ($''$) & $i$ ($^{\circ}$) & PA ($^{\circ}$)\\
		\hline
		J & 0.12$\pm$0.01 & $0.74^{+0.01}_{-0.01}$ & $0.19^{+0.01}_{-0.01}$ & $-0.009^{+0.001}_{-0.001}$ & $89.43^{+0.03}_{-0.03}$ & $165.06^{+0.01}_{-0.01}$\\
		H & 0.12$\pm$0.01 & $0.83^{+0.01}_{-0.01}$ & $-0.04^{+0.01}_{-0.01}$ & $-0.002^{+0.001}_{-0.001}$ & $88.71^{+0.08}_{-0.06}$ & $165.67^{+0.03}_{-0.03}$\\
		K1 & 0.14$\pm$0.01 & $0.87^{+0.07}_{-0.04}$ & $0.06^{+0.07}_{-0.04}$ & $0.004^{+0.003}_{-0.003}$ & $87.95^{+0.23}_{-0.23}$ & $166.12^{+0.15}_{-0.18}$\\
		\hline
		\hline
	\end{tabular*}
\end{table*}

\subsection{Surface Brightness} \label{sbsection}
\begin{figure*}
	\caption{\label{sb} \textbf{Left:} The disk emission in polarized intensity as a function of separation from the star. The two vertical grey dashed lines represent the estimated location of the peak polarization on either side of the disk. \textbf{Right:} The disk emission in total intensity as a function of separation from the star. Disk orientation is the same as in Figure \ref{vert}.}
	\includegraphics[width=1\textwidth]{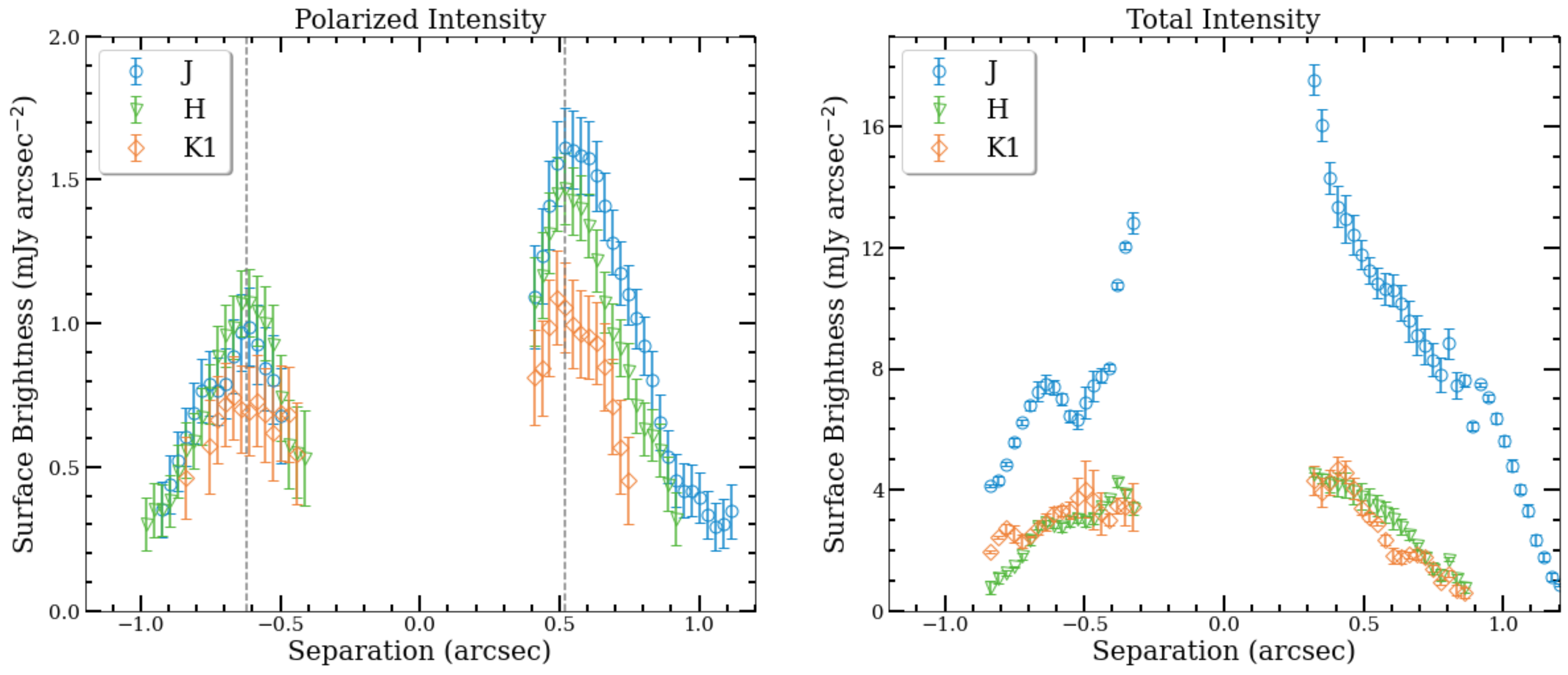}
\end{figure*}

Previous analysis of GPI observations for HD 111520 report a 2:1 brightness asymmetry in the $H$ band \citep{zd16b}. We measured the surface brightness of our multi-wavelength data to better constrain the brightness asymmetry across all three bands, as well as to measure the disk color. Here, the vertical offset profile is used to approximate the peak surface brightness location along the disk. The surface brightness is integrated along 2 by 12 pixel bins at a series of radii from the star, where 12 pixels is similar to the measured FWHM of the disk. For the polarized intensity, $1\sigma$ uncertainties are measured using noise maps derived from the $U_{\phi}$ data in each band. In contrast, 1$\sigma$ uncertainties for the total intensity are measured using the standard deviation of similar sized bins located at the same radial distance but outside the disk. We do not take into account the additional uncertainty in our total intensity data due to self subtraction, as we are mainly interested in the brightness asymmetry between the two extensions, but note that this uncertainty exists. Figure \ref{sb} shows the final surface brightness as a function of separation from the star for all three bands in both polarized and total intensity. Only data points with a S/N$\geqslant$3, are plotted.

By eye, a stark difference between the polarized and total intensity surface brightness profiles can be seen. While the total intensity appears to peak either within or near the focal plane mask, and then consistently decreases with distance, the polarized intensity instead peaks at a farther separation on either side. This difference can be easily explained by the scattering phase function of the disk, where the total phase function should peak at small scattering angles due to strong forward scattering \citep{milli2017}, while the polarized phase function should peak at larger scattering angles \citep{engler2017,milli2019}. Interestingly, the peak polarized intensity differs between the NW and SE extensions consistently between the three bands, where the NW extension peaks closer at $\sim0\farcs52$ from the star compared to the SE extension which peaks at a separation of $\sim0\farcs62$ from the star (corresponding to a difference of $\sim$11 AU). This suggests an eccentric disk with the NW extension located closer to the star compared to the SE extension.

Additionally, a clear brightness asymmetry can be seen in both polarized and total intensity. To probe the significance of this asymmetry, we compare the flux between the NW and SE extensions in all three bands. This is done by integrating the disk flux through two rectangular apertures (sized 30 by 15 pixels) placed on the NW and SE extensions at $0\farcs35$ to $0\farcs8$ from the star, and then comparing the total integrated flux between the two sides. Uncertainties for the polarized intensity are measured by integrating the flux of the noise maps over the same aperture, while for the total intensity the uncertainty is measured by taking the standard deviation of the same sized aperture located at the same radius but placed outside the disk. The NW/SE integrated flux ratio for each band, Pol and Spec, can be found in Table \ref{asymm_sum}. Our results show a similar trend between the polarized and total intensity with a large brightness asymmetry, most prominent in the $J$ band, of $\sim$1.8:1, although not as large as the 2:1 asymmetry reported in \cite{zd16b}. What is more surprising is the apparent wavelength dependency of the brightness asymmetry, where the asymmetry seen in the $H$ and $K1$ bands is significantly lower at only $\sim$1.5:1. We also find no brightness asymmetry at all in the $K1$-band total intensity, although this may be partially due to self-subtraction. Thus, the observed brightness asymmetry is strongest at shorter wavelengths and decreases as wavelength increases.

Through the surface brightness in each band, we can also extract the disk's color which can provide some information about its dust grain properties. This is because disk color is highly dependent on the scattering properties of dust grains which are affected by composition, porosity, and grain size. For example, a blue disk color at NIR wavelengths can be caused by sub-micron sized grains or very porous grains \citep{boccaletti03}. While it is difficult to disentangle these dust grain properties from the disk color alone, we would expect that the disk color should be the same across the disk given symmetrical dust grain properties. To measure the disk color, we use the same integrated flux used for measuring the brightness asymmetry. We then compare the integrated flux on both sides between each band, which are converted to magnitudes. Finally, given that these are scattered light observations, the stellar magnitude must be taken into account in order to measure the disk color. In this last step, the difference in magnitude of the star between each band is subtracted from the difference in disk magnitude between each band (ex. $J-H$ = $(J - H)_{disk} - (J - H)_{star})$. For the stellar magnitudes, we use the 2MASS $J$, $H$, and $K$ magnitudes (8.00$\pm$0.02 mag, 7.83$\pm$0.06 mag, and 7.72$\pm$0.02 mag respectively; \citealt{CR03}).

The derived disk colors can be found in Table \ref{color_sum}. We include measurements for both polarized and total intensity for comparison, but note again that the total intensity data likely suffer from severe self subtraction, and therefore provide less reliable measurements compared to the polarized intensity. These values show that the disk has a blue color at NIR wavelengths. However, what is even more interesting, is that the NW extension appears to be bluer than the SE extension in $J - H$ and $J - K1$, while the difference in color is still present but decreases significantly with $H - K1$ in both total and polarized intensity. This, along with the stronger asymmetry in the $J$ band, suggests we are probing different dust grain properties at shorter wavelengths, a possibility which will be discussed further in Section \ref{discussion}.

\begin{table}
	\centering
	\caption{\label{asymm_sum}Measured brightness asymmetry between the NW and SE extensions (NW/SE) with 1$\sigma$ uncertainties.}
	\begin{tabular}{cccc}
	    \hline
	    \hline
		Mode & J & H & K1\\
		\hline
		Pol & $1.77\pm0.07$ & $1.48\pm0.05$ & $1.45\pm0.08$\\
		Spec & $1.83\pm0.01$ & $1.44\pm0.01$ & $0.98\pm0.01$\\
		\hline
		\hline
	\end{tabular}
\end{table}

\begin{table}
	\centering
	\caption{\label{color_sum}Measured disk color using polarized and total intensity, with 1$\sigma$ uncertainties.}
	\begin{tabular}{ccc}
	    \hline
	    \hline
		Bands & NW Extension (mag) & SE extension (mag)\\
		\hline
		J-H (pol) & $-0.34\pm0.03$ & $-0.13\pm0.05$\\
		J-K1 (pol) & $-0.77\pm0.04$ & $-0.49\pm0.06$\\
		H-K1 (pol) & $-0.44\pm0.04$ & $-0.36\pm0.06$\\
		\hline
		J-H (spec) & $-1.59\pm0.01$ & $-1.17\pm0.01$\\
		J-K1 (spec) & $-1.77\pm0.01$ & $-1.16\pm0.01$\\
		H-K1 (spec) & $-0.18\pm0.01$ & $-0.01\pm0.01$\\
		\hline
		\hline
	\end{tabular}
\end{table}

\subsection{HST Observations} \label{sec:hst}
\begin{figure*}
	\caption{\label{fig:hst_offset} Vertical offset profiles of the GPI and HST data for HD 111520. Both images are rotated by $75\fdg7$ clockwise. The dark blue data points represents the GPI vertical offset in pol $H$ band, the light blue data points represents the HST vertical offset, while the orange data points represent the vertical offset of the bifurcation or `fork' feature observed with HST in the NW extension. The diagonal grey dashed lines are plotted to show the slopes of the SE and NW extensions of the disk halo seen with HST, while the horizontal grey dashed line shows the alignment of the GPI data. The vertical shaded regions at $1\farcs7$ show where the the disk halo changes from being relatively flat to sloped.}
	\includegraphics[width=\textwidth]{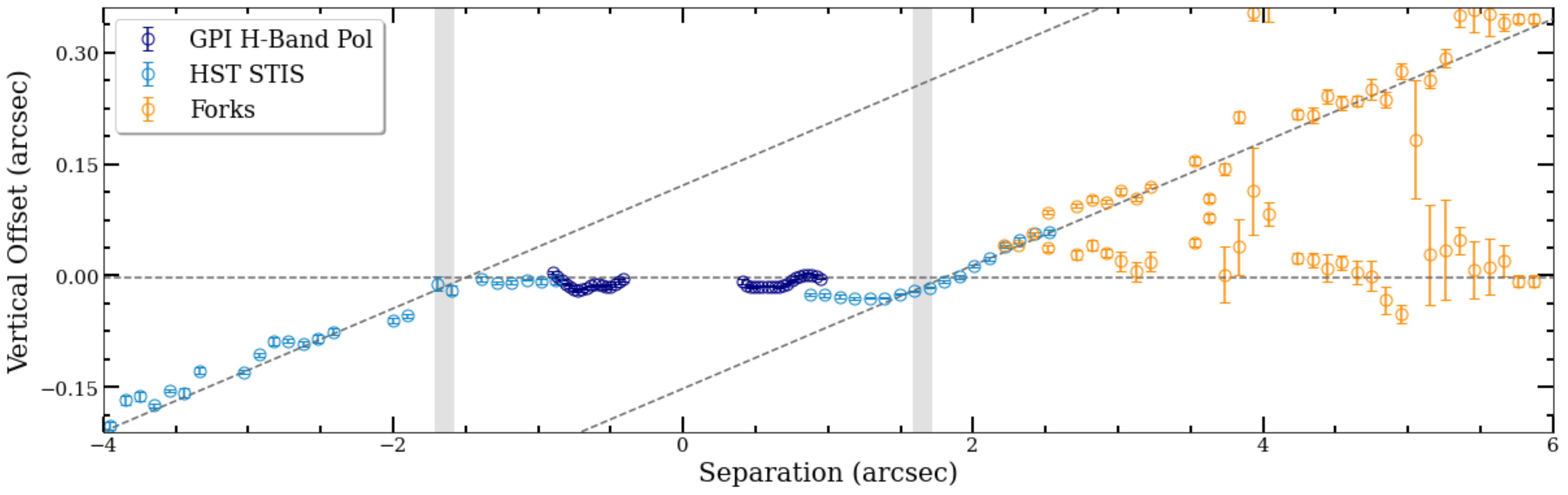}
\end{figure*}

While this paper is mainly focused on the micron sized dust grains of the debris disk around HD 111520 as observed with GPI, we also take a deeper look into observations with HST STIS. HST data are sensitive to the smallest dust grains set on highly eccentric parabolic orbits by radiation pressure that form the most extended component of debris disks, also known as the disk halo \citep{MH18}. These observations can be seen plotted in Figure \ref{hst}, with the GPI data in Spec $H$ band plotted in the center. Like the dust grains observed with GPI, the disk halo also features strong asymmetries. This includes an even larger brightness asymmetry of 5:1, a radial asymmetry between the two sides, as well as a bifurcation or `fork' feature observed in the NW extension \citep{DP15,zd16b}. To probe the disk halo structure further, we measure the vertical offset of the HST STIS image in a similar manner as in Section \ref{structure}, however, for the fork structure, we instead fit a double Gaussian profile instead of a single Gaussian profile. These results can be seen in Figure \ref{fig:hst_offset} along side the GPI vertical offset profile in the polarized $H$ band for comparison.

What we find is that the disk halo appears to be warped, where beyond radial separations of $\sim1\farcs7$, the vertical offset of the disk halo turns from being relatively flat and aligned with the GPI-imaged inner disk, to becoming sloped. While the SE extension warps downwards, the NW extension slopes upwards by roughly the same angle ($\sim3\fdg8$). We are also able to resolve the vertical offset of the fork down to $\sim2\farcs5$ and out to 6$''$. The warp in the NW appears to align with the upper fork, while on the other hand, the GPI data appears to be aligned with the bottom fork. While there is a small uncertainty in the disk PA for both the GPI and HST STIS instruments of $\sim0\fdg1$ each, this uncertainty does not change the alignment between the GPI data and the lower fork seen in the disk halo.

These results help clarify the disk morphology and show that the disk is being perturbed at large radial distances from the star. A planet companion as the potential cause for the warped structure and asymmetries of the disk halo will be addressed in Section \ref{discussion}.

\section{Discussion} \label{discussion}
Through the analysis of HD 111520's disk structure, surface brightness and disk color at multiple wavelengths, we have measured the disk's geometry, brightness asymmetry and placed constraints on the dust grain properties. We find that the brightness and radial asymmetry is strongest in the $J$ band and that the disk color of the NW extension is relatively bluer than the SE extension in $J-H$ and $J-K1$, while becoming more comparably blue in $H-K1$. This shows a possible wavelength dependence on all of the disk's observed asymmetries. As the disk color heavily depends on the dust grain properties, a difference between the color of the two extensions suggests we are probing different dust grain properties, either with the minimum dust grain size, composition, porosity, or a combination of these. The brightness asymmetry may also be explained by the disk eccentricity or an asymmetry in the dust mass distribution. In this section, we explore different scenarios by which the asymmetries and features seen in HD 111520 could have been formed given our results from the analysis of the GPI data, as well as the morphology of the disk halo observed with HST.

\subsection{Disk Eccentricity}
While it is difficult to derive information about a possible disk offset along the major axis through the vertical structure, given the high inclination, there is another clue that points towards the disk being eccentric. This includes the polarized intensity surface brightness profile, where we measure the brightness in the NW extension to peak closer to the star compared to the SE extension. In the case where there is no disk eccentricity, we would expect that the surface brightness would peak in the same location on either side. If the disk instead has some eccentricity, this would cause one side of the disk to be closer to the star, bringing the peak surface brightness of that side inwards and creating a pericenter glow \citep{Wyatt99,Pan2016}. For HD 111520, this would mean that the NW extension lies closer to the star than the SE extension. If this is the case, this could at least partially explain the brightness asymmetry seen, as the NW extension would be receiving a higher amount of stellar light. 

Given the separations of the two polarized peaks ($\sim0\farcs52/0\farcs62$), we can place constraints on the possible eccentricity of the disk. Assuming that the argument of periapsis and the peak polarized intensity are both along the projected major axis, this leads to estimated eccentricity of 0.09. If the argument of periapsis is not along the projected major axis, the true eccentricity would be even larger making 0.09 a lower limit, however, this lower limit may be slightly overestimated if the peak polarized intensity is not along the projected major axis. Even so, such an estimated eccentricity is significant, and only observed in a handful of debris disks. One exciting aspect of these measurements is that an eccentric disk may point towards a hidden planetary companion, as planets on eccentric orbits have been shown to induce eccentric disks through dynamical simulations \citep{Lee:2016aa,Lin2019}, as well as observations (e.g. Fomalhaut; \citealt{PK2005}, \citealt{MM17}). However, further analysis is required to see if an eccentric disk can fully explain the brightness and structural asymmetries observed.

As the surface brightness is related to 1/$r^{2}$, with $r$ being the radial separation from the star, we can also estimate what the brightness asymmetry should be given the different locations in the projected peak surface brightness. Taking the ratio of 1/$r^{2}$ between the two sides, we find that a brightness asymmetry of $\sim$1.3:1 would be expected at a distance of between $0\farcs5$ and $0\farcs6$ from the star, assuming that the projected peak polarized intensity is occurring at the disk ansae. If in fact the projected peak polarized intensity is occurring elsewhere along the projected major axis, the expected brightness asymmetry would be even less. This is considerably lower than the 1.5:1 to 1.8:1 brightness asymmetry observed in the polarized intensity at these distances, meaning that while an eccentric disk may partially explain the observed brightness asymmetry, there must be another mechanism at play. This is further supported by the fact that, in the $J$ band, the NW extension is much more extended than the SE extension, which should not be the case for an eccentric disk with a stellar offset towards the NW. Additionally, an eccentric disk alone would not explain the differences in brightness asymmetries we observe between bands, as well as the relatively bluer disk color observed in the NW extension. These inconsistencies again show that we need another phenomenon to help explain the disk's asymmetries. 

\begin{figure*}
	\caption{\label{hst} HD 111520 as seen by HST STIS (0.59 $\mu$m) and GPI (1.65 $\mu$m), both rotated by $75\fdg7$ clockwise. The HST image is viewed in log scale. The southern extension is significantly dimmer than the northern extension, with brightness asymmetries of 1.5:1-1.8:1 within GPI's FOV and 5:1 in HST's FOV \citep{DP15}. From GPI to HST scales, it is clear that there is a large-scale change in the disk. The South to North extension of the disk appears to have an asymmetrical geometry, with a possible bifurcation, or `fork', seen on the Northern side \citep{DP15}. The central arrow represents the direction of the proper motion for the system.}
	\includegraphics[width=\textwidth]{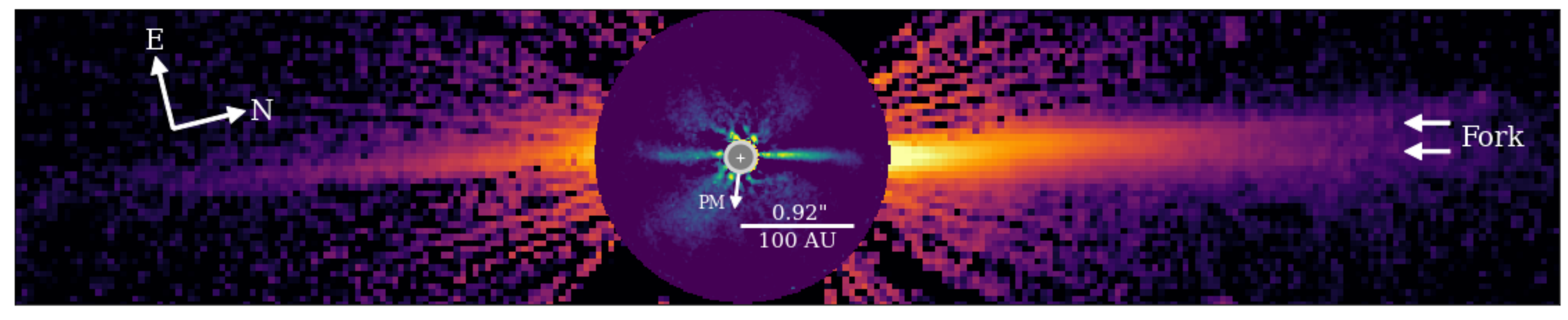}
\end{figure*}

\subsection{ISM interaction}
Another possible explanation for the observed asymmetries is an interaction with the ISM. By passing through a dense ISM cloud, this can cause small dust grains in particular to be pushed towards the side of the disk opposite the direction of motion \citep{DJ09}. This redistribution of small grains from one side of the disk to the other can cause one disk extension to become relatively brighter and more extended at shorter wavelengths, both of which we observe for HD 111520 with GPI. Similar features can be seen with HST, where the NW extension is brighter and more extended than the SE extension, as shown in Figure \ref{hst}. If the NW extension is populated with smaller grains compared to the SE extension, this would also explain the much bluer disk color in the NW extension observed in $J$-$H$ and $J$-$K1$. However, for this to occur, the proper motion of the star must align in the correct direction. In the case of HD 111520, to create a NW extension that is brighter and bluer than the SE extension through an ISM interaction, the proper motion of the system must also be pointing towards the SE direction. The current measurements for HD 111520's proper motion are -35.3 mas/yr in RA and -16.7 mas/yr in Dec \citep{gaia21}. Based on these measurements, this would mean that the proper motion is pointing towards the SW direction rather than the SE (see Figure \ref{hst}), essentially ruling out an ISM interaction as the cause of the disk's color, radial and brightness asymmetries.

\subsection{Giant Impact}
Another possible explanation of the features of HD 111520's debris disk would be a giant impact of two large bodies within the disk. Such an impact would generate an avalanche of small dust grains that would start out in a clump at the collision site, and would be redistributed throughout the disk over time. This scenario is consistent with the large brightness asymmetry observed and the difference in disk color, assuming that the impact occurred in the NW extension. A giant impact would have had to occur recently, on the order of a few thousand orbits (1 Myr at 50 AU), in order for the pinched collision point to still be present, causing the observed asymmetries \citep{AJ14}. Because all particles are forced to pass through the collision point, this allows for further collisions and generation of small dust grains. After the collision point is smeared out, the asymmetry in the generated dust grains would start to become washed out due to collisional diffusion and blown out due to radiation pressure, returning the disk to an axisymmetric state over time. There are a few issues with this scenario, however. One issue is that typically the side opposite of the impact would be more radially extended \citep{AJ14}, whereas for HD 111520 we observe the opposite, although it is important to note that radiation pressure would also be playing a role in blowing out the small grains and we may not be sensitive to dust grains on the radially extended side, given the expected lower surface brightness. Another issue is that verification of this hypothesis is unfortunately not possible with GPI data alone.

Further evidence would include analyzing the dust mass distribution, which can be obtained through mm observations, as the majority of the observable disk mass lies within mm-sized dust grains and these grains are less susceptible to radiative forces in the disk. If the dust mass distribution in the mm-sized dust grains is also asymmetric, this would support a giant impact scenario and would help explain the very large surface brightness asymmetry. Higher resolution CO observations would also be extremely useful in this case, as a large impact may release a large amount of CO gas. If a concentration of CO is observed in the NW extension, this would help confirm the possibility of a giant impact as it has for the $\beta$ Pic debris disk \citep{WD14}. While a marginally-resolved continuum observation of the HD 111520 disk has been made with ALMA at 1240$\mu$m \citep{LS16}, establishing an asymmetric dust mass distribution requires the disk to be well resolved. Along with the continuum, low resolution CO observations were also taken but no CO was detected. It is worth noting that there are several disks that are asymmetric in scattered light with symmetric mm-sized grains (HR 4796, HD 61005; \citealt{Olofsson19,buenzli10,macgregor18}), however, such observations would still provide important information about the overall disk morphology. If there is an asymmetry in the mm-sized grains, then a local enhancement due to a large impact is likely, while symmetric mm-sized grains would support the need for an alternative mechanism to explain the asymmetries present in the scattered light in addition to an eccentric disk. Thus, high resolution ALMA observations of the mm continuum emission will be necessary in order to support or refute the hypothesis of a recent giant impact.

\subsection{Disk Halo}
In Section \ref{sec:hst}, our analysis of the HST observations revealed that the disk halo is warped past $\sim1\farcs7$ on either side of the disk by $3\fdg8$. Additionally, the fork structure in the NW extension is resolvable from $2\farcs5$-$6''$. While the upper part of the fork is aligns with the warp of $3\fdg8$ in the NW extension, the lower fork is aligned with the inner part of the disk as observed with GPI. This alignment can rule out certain scenarios for the cause of the fork structure, such as self shadowing from a higher dust mass in the NW extension as suggested in \citep{zd16b}, which would require the larger grains to be aligned between the two forks.

One possible explanation for the fork's existence is that somewhere in the system exists an undetected planetary companion with a mutual inclination relative to the disk's. Such a planet has been shown to cause an ``X" shape in the disk morphology in the dust density distribution, which may be observed as a fork like structure on either one or both sides of the disk depending on the viewing angle \citep{Pearce2014}. However, this mainly applies to larger grains while smaller grains are also affected by radiation pressure, which may lead to a diminished effect. Such a planet may also cause the disk to become warped, such as the case with the inner disk of $\beta$ Pic \citep{mouillet97,Dawson2011,DA15}. In the case of HD 111520, the warp observed would suggest a planet inclined at $\sim3\fdg8$ relative to the disk, orbiting at a distance greater than $1\farcs7$ ($\sim$184 AU) where the warp is observed. While dynamical modelling would be needed to place better constraints on a planet perturber, the complex structure of the disk halo does show that a planet companion at large separations likely exists in this system.

\section{Conclusion}

Using deep, multi-wavelength GPI data of HD 111520's debris disk in both polarized and total intensity, we have been able to measure the disk's vertical structure, surface brightness profiles, and disk color.

\begin{itemize}
    \item We find that the disk has an intrinsic FWHM $0\farcs12$-$0\farcs15$ between the $J$, $H$ and $K1$ bands. The profile also exhibits a positive trend between the FWHM and radial distance in the NW extension, while in the SE extension a FWHM enhancement is observed at $\sim0\farcs5$ which then flattens out past $0\farcs6$.
    \item Measuring the vertical offset along the disk, we find that the west side is the front side of the disk. The measured disk spine also lies within $0\farcs03$ of the star location, showing the disk to have an inclination close to (but not quite) 90$^{\circ}$. This is confirmed through modeling the vertical offset profile, where we derive an inclination roughly between 87$^{\circ}$-89$^{\circ}$ and a PA of 165$^{\circ}$-166$^{\circ}$ from the best fitting models. We also derive disk offsets of $\delta_{x}$ and $\delta_{y}$, however, the disk is too highly inclined to place a good constraint on the disk offset along the major-axis. No disk offset is found along the minor-axis.
    \item Through characterization of the disk structure, we find a radial asymmetry exists, with a NW side that is more extended than the SE. However, this radial asymmetry is only present in the $J$ band, while not present in the $H$ and $K1$ bands.
    \item By measuring the surface brightness in all three bands, we find that the polarized intensity and total intensity have two very different profiles. Additionally, the polarized intensity peaks closer to the star in the NW extension compared to the SE extension, suggesting that the disk has an eccentricity of $\gtrsim$0.09. Although given an offset of $\sim$11 AU along the projected major axis, this eccentricity would not be sufficient enough to be the sole cause of the disk's brightness asymmetry.
    \item Comparing the surface brightness on either side shows a 1.5:1-1.8:1 brightness asymmetry, slightly less than what is observed in \cite{zd16b}. This asymmetry appears strongest in the $J$ band and decreases with wavelength.
    \item Similarly, measuring the disk color between each band in polarized intensity shows that the NW extension is relatively bluer than the SE extension in $J-H$ and $J-K1$, while this trend is not as strong in the $H-K1$ color.
    \item Through measuring the vertical offset of the disk halo as seen with HST, we find that the small grains at large separations ($>1\farcs7$) are highly warped on either side. We also are able to measure the fork down to $\sim2\farcs5$, where the GPI data appears to align with the lower fork.
\end{itemize}
  
Given the possibility of an eccentric disk, as well as the warped morphology of the disk halo, this suggests that there may be at least one planet perturber in this system, although dynamical modelling is needed to test this scenario. On the other hand, one way in which we can explain the large brightness asymmetry, as well as the difference in disk color between the two extensions, is if the two extensions contain differing dust grain properties, such as smaller grains in the NW extension. This can be caused by two scenarios: 1. An interaction with the ISM, or 2. a giant impact. While an ISM interaction would provide a straight-forward explanation, the proper motion is not in the correct direction to account for the asymmetries between the NW and SE extensions. On the other hand, a recent giant impact between two large bodies may also be the source for the surface brightness and disk color asymmetries; however, further observations/evidence are needed to probe this scenario further. Specifically, a sub-arcsecond resolution image of the dust mass distribution and CO through higher resolution ALMA observations is essential.

HD 111520 serves as a very interesting and unique system to study, with such a large brightness asymmetry compared with other debris disks, as well as a complicated overall disk morphology. By studying the disk in greater detail, we can gain a greater understanding of the ways in which dynamical perturbations and collisions can affect disk morphologies, as well as how these types of debris disk systems evolve.

\acknowledgements{The authors wish to thank the anonymous referee for helpful suggestions that improved this manuscript. This work is based on observations obtained at the Gemini Observatory, which is operated by the Association of Universities for Research in Astronomy, Inc. (AURA), under a cooperative agreement with the National Science Foundation (NSF) on behalf of the Gemini partnership: the NSF (United States), the National Research Council (Canada), CONICYT (Chile), Ministerio de Ciencia, Tecnolog\'{i}a e Innovaci\'{o}n Productiva (Argentina), and Minist\'{e}rio da Ci\^{e}ncia, Tecnologia e Inova\c{c}\~{a}o (Brazil). This work made use of data from the European Space Agency mission \emph{Gaia} (\url{https://www.cosmos.esa.int/gaia}), processed by the \emph{Gaia} Data Processing and Analysis Consortium (DPAC, \url{https://www.cosmos.esa.int/web/gaia/dpac/consortium}). Funding for the DPAC has been provided by national institutions, in particular the institutions participating in the Gaia Multilateral Agreement. This research made use of the SIMBAD and VizieR databases, operated at CDS, Strasbourg, France. We thank support from NSF AST-1518332, NASA NNX15AC89G and NNX15AD95G/NEXSS. KAC and BCM acknowledge a Discovery Grant from the Natural Science and Engineering Research Council of Canada.}

\vspace{5mm}
\facilities{Gemini:South}

\software{pyKLIP (\citealt{JW15}),
Gemini Planet Imager Pipeline (\citealt{per14}, \url{http://ascl.net/1411.018)},
emcee (\citealt{df13}, \url{http://ascl.net/1303.002}),
corner (\citealt{corner}, \url{http://ascl.net/1702.002}), 
matplotlib (\citealt{Hunter07}; \citealt{droettboom17}), 
iPython (\citealt{perez07}), 
Astropy (\citealt{Collaboration:2018ab}),
NumPy (\citealt{Oliphant_06}; \url{https://numpy.org}),
SciPy (\citealt{Virtanen_20}; \url{http://www.scipy.org/})}

\clearpage

\bibliographystyle{aasjournal}
\bibliography{ms.bib}

\listofchanges
\end{document}